\documentclass[final]{agujournal2019}
\usepackage{url} 
\usepackage{soul}
\usepackage{aas_macros}
\usepackage{amsmath}

\draftfalse
\journalname{Space Weather}

\begin{document}

%
%
\twocolumn[
\begin{@twocolumnfalse}
 
\title{Detection and Characterisation of a Coronal Mass Ejection using Interplanetary Scintillation measurements from the Murchison Widefield Array}

%
%




\authors{J. Morgan\affil{1}, P. I. McCauley\affil{2}, A. Waszewski\affil{1}, R. Ekers\affil{3,1}, R. Chhetri\affil{1,4}}
\affiliation{1}{International Centre for Radio Astronomy Research, Curtin University, 1 Turner Avenue, Bentley, WA 6102, Australia}
\affiliation{2}{School of Physics, University of Sydney, Sydney, NSW 2006, Australia}
\affiliation{3}{CSIRO Space and Astronomy, P.O. Box 76, Epping, NSW 1710, Australia}
\affiliation{4}{CSIRO Space and Astronomy, P.O. Box 1130, Bentley, WA 6102, Australia}




\correspondingauthor{J. Morgan}{john.morgan@icrar.org}



\begin{keypoints}
\item We are able to recover a Coronal Mass Ejection detected in coronagraph images in Interplanetary Scintillation (IPS) observations taken 33 hours later.
\item The unprecedented number of lines of sight in our IPS observations allow us to image the CME and localise it to degree-level accuracy.
\item The CME's location in the IPS observation is consistent with fast-CME propagation, and confirms a `broad-side' (plane of sky) trajectory.
\end{keypoints}

%
%

%
%


\begin{abstract}
We have shown previously that the Murchison Widefield Array (MWA), can detect hundreds of Interplanetary Scintillation (IPS) sources simultaneously across a field of view $\sim30^\circ$ in extent.
To test if we can use this capability to track heliospheric structures, we undertook a search of 88 hours of MWA IPS data, and identified an observation likely to have a significant Coronal Mass Ejection (CME) in the field of view.
We demonstrate that in a single 5-minute MWA observation we are able to localise and image a CME $\sim$33 hours after launch at an elongation of $\sim37^\circ$ from the Sun.
We use IPS observables to constrain the kinematics of the CME, and describe how MWA IPS observations can be used in the future to make a unique contribution to heliospheric modelling efforts.
\end{abstract}
\end{@twocolumnfalse}
]
\section{Introduction}
Interplanetary Scintillation (IPS) is a phenomenon discovered by \citeA{Clarke:1964} that arises when the turbulent solar wind crosses the line of sight to distant, compact radio sources.
As a result, the scintillation signature encodes information on the radio source \cite{1964Natur.203.1214H}, and the turbulent medium responsible for the scintillation.
Early on, IPS was used to track structures out through the heliosphere \cite{1968PASA....1..142D}; many decades before the role of Coronal Mass Ejections (CMEs) in connecting events on the Sun to Space Weather in the near-Earth environment was widely recognised \cite{1993JGR....9818937G}.
IPS can also provide velocity measurements of the solar wind; either by the use of multi-station IPS \cite{1967Natur.213..343D} or via fitting the power spectrum measured at a single site.
\cite{1971ApJ...168..543Y,1990MNRAS.244..691M}.
IPS continues to be used for monitoring of the Solar Wind \cite<see, e.g.>[and references therein]{2020FrASS...7...76J,2021ApJ...922...73T}.

\citeA{2018MNRAS.473.2965M} adapted the IPS technique to exploit the capabilities of widefield interferometers such as the Murchison Widefield Array \cite<MWA;>{2013PASA...30....7T}, allowing hundreds of IPS sources to be measured simultaneously.
Here, we aim to show how this unprecedented number of lines of sight can be used to image structures in the heliosphere.
We present the first MWA IPS detection of a CME in the inner heliosphere \cite<c.f.>[a nightside detection]{2015ApJ...809L..12K}.
This observation was selected for analysis from over 1000 observations \cite{2019PASA...36....2M} as a good candidate for containing a CME based on a previous coronagraph detection.

The paper is organised as follows:
in Sect.~\ref{sec:method} we describe how we chose a target observation and used it to map interplanetary turbulence.
In Sect.~\ref{sec:ancillary}, we describe the CME at launch time, including an analysis of solar coronagraph data to determine a plane-of-sky velocity for the CME after its acceleration.
In Sect.~\ref{sec:results} we describe the results our of MWA observation, and compare it with near-contemporaneous IPS data from the multi-station ISEE IPS array.
In Sect.~\ref{sec:analysis} we analyse these results, and in Sect.~\ref{sec:discussion} we discuss our findings and future plans.

\section{Method}
\label{sec:method}
\subsection{Coordinate systems}
For the trajectory of a CME we use a Sun-centred spherical coordinate system where $\theta$ refers to the angle with the plane-of-sky (i.e. positive $\theta$ is towards Earth, whereas $\theta=0$ denotes a trajectory normal to the Earth-Sun line) and $\phi$ is the angle in the plane-of-sky, measured from solar west through north \cite<see>[for a more rigorous description]{2004JGRA..109.3109X}.
We also use Earth-based observer-centred coordinates, where $\epsilon$ is used for solar elongation, and $D$ for the distance from the Earth.
$\epsilon$ and $\phi$ together form a heliocentric polar coordinate system for describing the location of any point on the sky relative to the Sun (as observed from Earth).
A relationship between $\epsilon$, $\theta$ and $D$ can be determined via the sine rule:
\begin{equation}
    \label{eqn:sinerule}
    D\left(\textrm{AU}\right) = \frac{\sin{\theta'}}{\sin{\left(\theta'+\epsilon\right)}} 
\end{equation}
where $\theta'=\theta-\pi/2$, with the simplifying assumption that the Sun-Earth distance is 1\,AU.
\subsection{Identification of main target field.}
\label{sec:xmatch}
In order to identify MWA IPS observations that were likely to contain a CME, we began with a list of 1062 observations taken between 2015-12-23 and 2016-08-02, described in detail in \citeA{2019PASA...36....2M}.
The observations used here cover 154.24\,MHz--169.6\,MHz contiguously, and are 576\,s in duration.
Essentially, observations were made of fields offset 30$^\circ$ from the Sun, at various position angles.
The MWA has a nominal Field of View (FoV) 30$^\circ$ across, giving access to solar elongations of 15$^\circ$--45$^\circ$.
This is a near optimal range for IPS observations at 162\,MHz (an observing frequency which we have chosen to provide a good balance of sensitivity and field of view), and also places the Sun in the null of the instrumental response, mitigating its impact.

For a list of CME events that might be detectable in an MWA IPS observation, we used the catalogue generated by the Computer Aided CME Tracking algorithm \cite<CACTus;>{2004A&A...425.1097R} using images from the Large Angle Spectral Coronagraph  \cite<LASCO;>{1995SoPh..162..357B}.
This publically available catalogue is described by \citeA{2009ApJ...691.1222R}, and provides a launch time, plane-of-sky velocity, opening angle, and a position angle for each CME.
Assuming that all of these CMEs continued to propagate at the velocity determined by CACTus, and utilising the opening angle, we were then able to determine which CMEs would have some overlap with the MWA FoV (note that this analysis predates the publication of  CMEChaser \cite{2020SoPh..295..136S} which takes a broadly similar approach).
This method is not intended to be highly accurate, but is a simple and practical way for us to triage our data in an automated fashion. It is nonetheless sufficient for our purposes since with the wide field of view of the MWA, we have a good chance of capturing the CME, even with an error approaching $\pm50$\%.

These very broad criteria resulted in 417 matches between observations and CMEs.
Of these, the most promising matches were 21 for which the projected centroid position was within the MWA FoV, and the opening angle covered the entire MWA FoV.
This was a small enough number that each crossmatch could be inspected manually.
A candidate was chosen on the basis of a predicted location of the CME close to the centre of the FoV, and an unambiguous and strong detection in LASCO difference images.
The CACTus CME corresponding to our chosen MWA observation is number 0045 in the CACTus ``quick look'' catalogue for 2016-05, with a launch time of 2016-05-15T15:24 (see
Figure~\ref{fig:lasco}).

\subsection{The target observation}
\label{sec:target}
The corresponding MWA observation began at 2016-05-17T00:25:35 UTC, 33 hours after the CME launch time.
Calibration, imaging, and determination of the scintillation index of each IPS sources was then carried out exactly as described in \citeA{2018MNRAS.473.2965M}.
We note that due to the unique capabilities of the MWA, the methodology used (in particular, measurement of the thermal noise, and derivation of the scintillation index) differs considerably from that used at other IPS observatories.
The method is summarised below, and the reader is referred to \citeA{2018MNRAS.473.2965M} for further details.

Briefly, a ``standard image'' was made using the full observation, using standard interferometry software, WSCLEAN; \cite{2014MNRAS.444..606O,2017MNRAS.471..301O}.
Images had $2400\times1$\,arcmin pixels in both dimensions, and a uniform visibility weighting scheme was used to maximise image fidelity and resolution.
Next, individual snapshot images were made of each 0.5-s observing interval with the same image dimensions, but with a natural weighting scheme to maximise sensitivity.
A filter with a bandpass of 0.1\,Hz--0.5\,Hz was then applied to the timeseries corresponding to each pixel.
This has the effect of emphasising the IPS signal relative to the noise while eliminating variability due to ionospheric effects.
The standard deviation of these filtered timeseries (one per pixel) gives the ``variability image'', from which can be derived the variance due to IPS alone, with instrumental white noise being measured and subtracted using the majority of pixels which do not contain an IPS source.
The noise-subtracted measurement of the brightness of each source in the variability image, along with its brightness in the ``standard image'', provides the numerator and denominator of the scintillation index.

\subsection{The reference observation}
Changes in solar wind density along a line of sight to an IPS source causes a change in the scintillation index relative to a baseline level: the so-called `g-level' \cite<e.g.>{1982Natur.296..633G}.
This g-level can then be mapped to a particular density \cite<e.g.>{1986P&SS...34...93T}, though this is not directly relevant to the current work.
Since the baseline scintillation level of a particular source cannot be known \textit{a priori}, it is typically determined over a number of days in order to average over stochastic fluctuations.

Most of the IPS sources detected in our target observation are not known IPS sources, and so their typical scintillation level cannot be known \textit{a priori}.
For this initial demonstration, we adopt the simple approach of using just a single observation of the same field from 24 hours previously; thus, the g-level for a particular source is simply the ratio of the scintillation index (calculated as described in Sect.~\ref{sec:target}) between the target and reference observations.
We do not expect the reference observation to contain a CME according to the cross-match described in Sect.~\ref{sec:xmatch}.
This does not preclude structure which escaped classification by CACTus; however it is unlikely that an event as significant as ours would be missed.

Balancing the risk of other heliospheric transients contaminating our reference observation, this approach has several advantages.
First, the Sun has only moved $\sim1^\circ$ relative to our IPS radio sources, a negligible change in elongation.
Furthermore, slower variations in the heliosphere, such as the locations of fast and slow streams, will remain roughly constant over 24 hours, and so our calculated g-levels will reflect only the transient changes in the heliosphere which we wish to detect.

\section{Ancillary data}
\label{sec:ancillary}
In order to characterise the CME as fully as possible, we identified relevant events in the Solar and Geophysical event list published by the Space Weather Prediction Center (SWPC). 
We also acquired Geostationary Operational Environmental Satellites (GOES) soft X-ray timeseries data and LASCO images. 
Contemporaneous IPS data from the ISEE IPS array in Japan is described in Sect.~\ref{sec:isee}.

\begin{figure*}
	\center
	\includegraphics[width=0.9\textwidth]{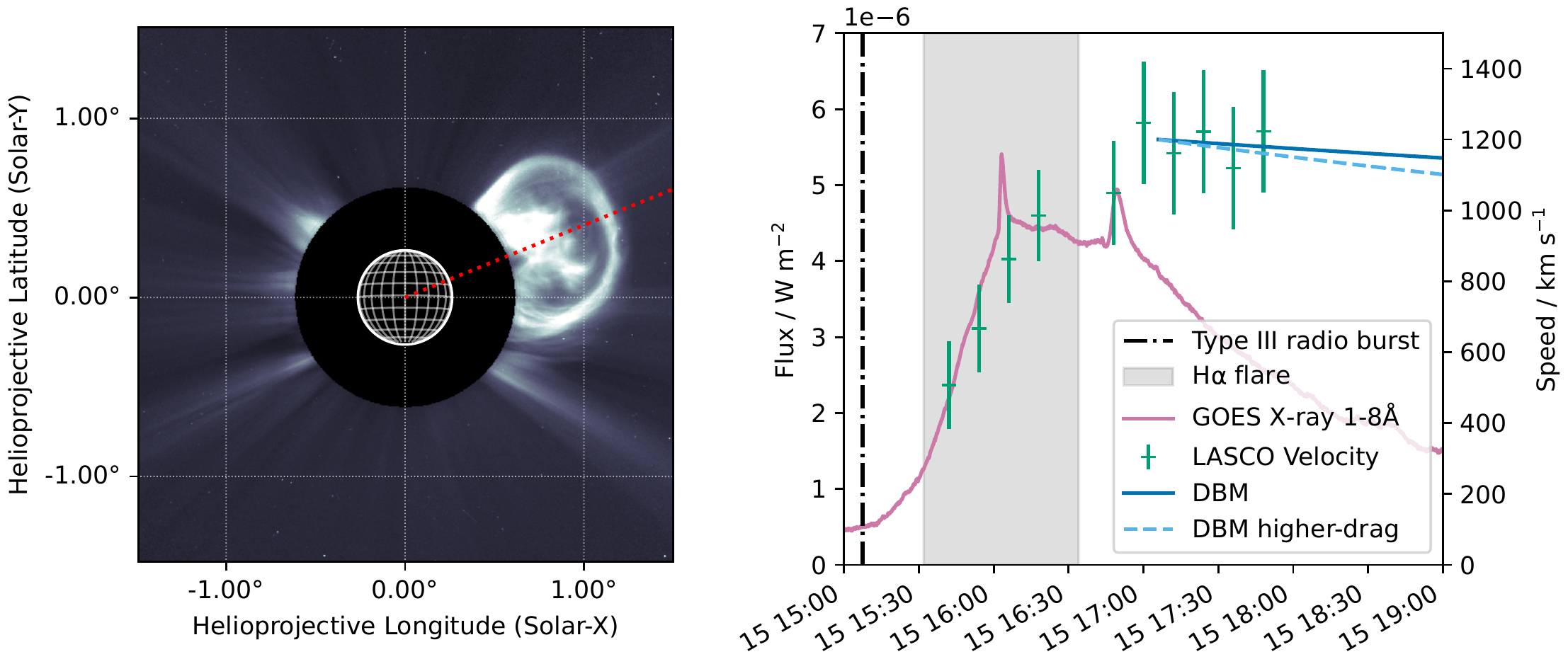}
	\caption{\label{fig:lasco} Left panel: LASCO C2 image of the CME at 2016-05-15T16:12:06.
     Red dotted line indicates approximate axis of symmetry of CME, estimated from this figure.
     Right panel: CME velocity overlaid on X-ray flux.
     Dash-dot line indicates time of Type-III radio burst.
     Grey range indicates H-alpha flare.
     Each green velocity point is based on a pair of measurements of CME front location separated in time.
     Error bars are based on 20'' error in C2 and 40'' error in C3.
     Blue lines show velocity modelled using DBM (assuming plane of sky; see Sect.~\ref{sec:analysis}).
     Velocity and X-ray flux axes are scaled arbitrarily.} 
\end{figure*}

The relevant findings are summarised in Figure~\ref{fig:lasco}.
The SWPC event list recorded a Type III radio burst, a C-class X-ray flare and an H$\mathrm{\alpha}$ solar flare, all of which are likely related to the CME.
Importantly, the H$\mathrm{\alpha}$ flare is localised to heliolatitude 10$^\circ$ N, heliolongitude 62$^\circ$ W.
This corresponds to a radial trajectory $\phi=$10$^\circ$ (north of West), $\theta=+28^\circ$ (from the plane of sky towards the Earth).

While the CACTus CME parameters were sufficient for the cross-matching process described in Sect.~\ref{sec:xmatch}, we decided to cross-check by performing our own analysis on the original coronagraph images.
The CME leading edge was well-defined in 5 C2 images and 7 C3 images (the left panel of Figure~\ref{fig:lasco} shows one C2 image).
We manually measured the location of the CME front along this line, and the velocity implied by each pair of consecutive height measurements is shown in the right panel of Figure~\ref{fig:lasco}.

The acceleration and match of the velocity and X-ray profiles are typical for a fast CME \cite<e.g. the ``intermediate acceleration'' CME studied by>{2004ApJ...604..420Z}.
Although the exact dynamics are unclear (and not relevant here), it reaches a final speed of 1\,200 km s$^{-1}$ at 2016-05-15T17:06, at which point it has an elongation of (2.56$\pm$0.03)$^\circ$.
After this point it continues at near-constant speed, possibly slowly decelerating at a rate consistent with the `Drag Based Model' of \citeA{2013SoPh..285..295V}.


\section{Results}
\label{sec:results}
\begin{figure*}
    \centering
    \includegraphics[width=0.7\textwidth]{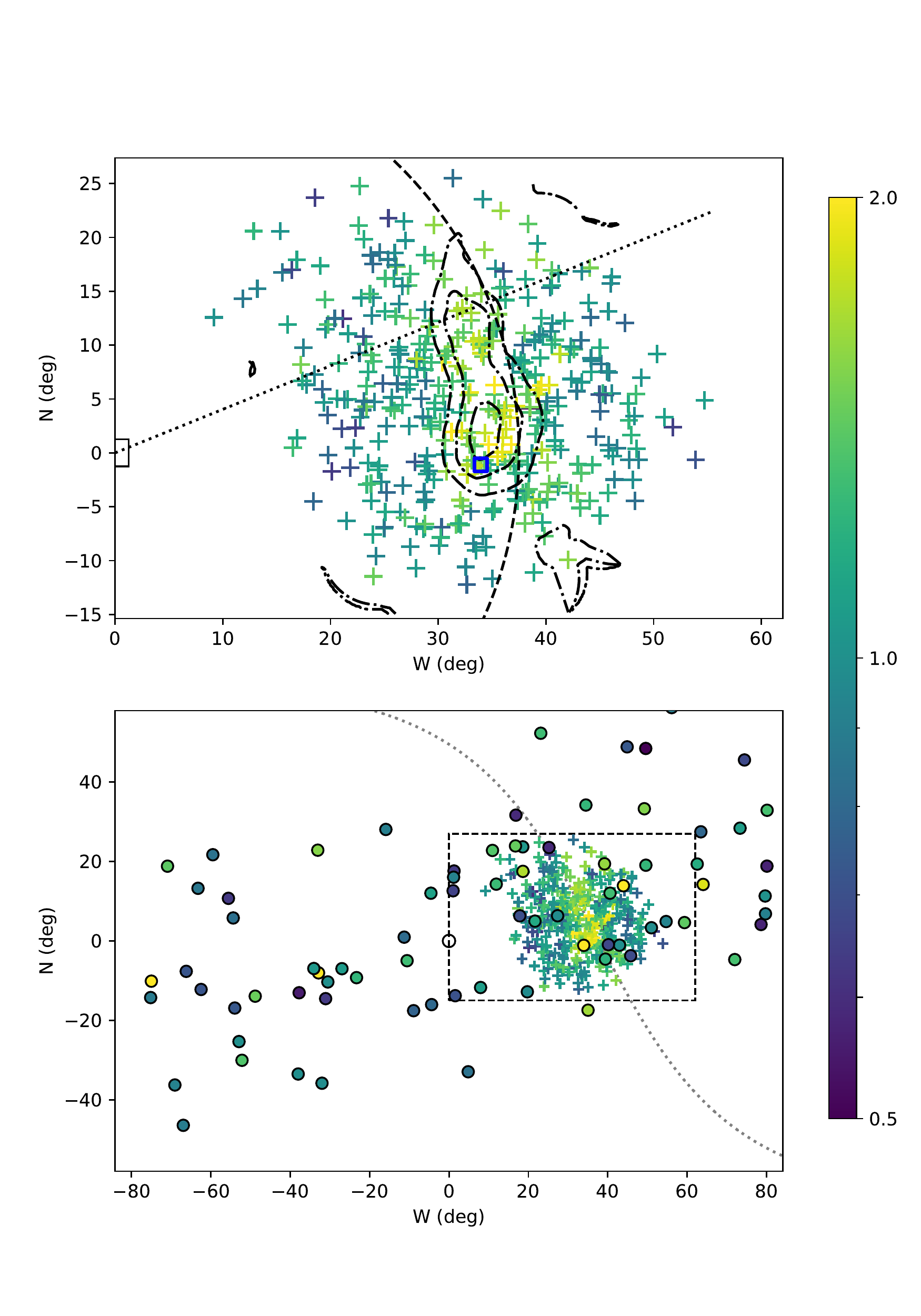}
    \caption{\label{fig:skymap}
    IPS observations in helioprojective coordinates.
    Top panel: sky map of g values derived from MWA data. 
    Dotted line indicates direction of CME measured in LASCO images (i.e. red line in left panel of Figure~\ref{fig:lasco}).
    Dashed line indicates approximate location of the CME front (37.5$^\circ$) estimated from this figure.
    Dash-dot lines are contours of smoothed g-value at 1.4, 1.5 and 1.6.
    Blue open square indicates location of ISEE source.
    Bottom panel: sky map with MWA g-levels (crosses) and ISEE g-levels (filled circles).
    Dotted line indicates location of ISEE meridian at the time of the MWA observation.
    Dashed rectangle indicates the boundary of the top panel, with the position of the Sun indicated by the open circle.}
\end{figure*}

Once both standard and variability images had been produced for both observations, we are left with the common problem in astronomical imaging of identifying significant detections within our noisy images.
For this, we used the widely used \textsc{Aegean} software package \cite{2018PASA...35...11H} to identify sources in all 4 images.
In all, 397 IPS sources were detected at 5$\sigma$ in both variability and continuum in both observations.
The g-level for each of these 397 sources was determined by computing the ratio of scintillation index between the target and reference observation.
Figure~\ref{fig:skymap} shows the g-levels obtained for our observation.
The most obvious feature is a band of enhanced g-levels in the solar elongation range 32$^\circ$--37$^\circ$.
These g-levels range from 1.3 to 1.9.
In contrast, areas outside this area of enhancement have g-levels much closer to 1, except for occasional outliers.

We caution that at elongations $<20^\circ$ the IPS regime for the background solar wind is tending towards strong scintillation, which is associated with a saturation of the scintillation index at 1.0, thus the g-level metric is not sensitive to heliospheric structures closest to the Sun.
However, for elongations beyond approximately 27$^\circ$, sources can be expected to show a range of g levels from 0.5--2.0, so this should not affect our primary result.

As an alternative way to present these data, we also used these point measurements of the g-level to construct a `g-map' using a Radial Basis Function (RBF), with the weight of each source being determined by the inverse square of distance \cite{Epanechnikov:1969}
\begin{equation}
    w\left(r\right) = 
    \begin{cases}
    1-\left(\frac{r}{5}\right)^2 &;\ r<5^\circ \\
    0 &;\ \text{otherwise} \\
    \end{cases}
\end{equation}
where $r$ is the distance of each source from the point of interest. 
Three contours of this g-map are shown in the top panel of Figure~\ref{fig:skymap}, to further illustrate the approximate extent of the area of enhanced scintillation.

The very high density of points (approximately 1 per square degree) in the MWA allow us to constrain the location and demarcate its extent.
The region is extended tangentially to the Sun, with the bulk of the feature, and the highest g-values, lying below the centre line of the CME in the coronagraph images

\subsection{Comparison with ISEE data}
\label{sec:isee}
In the bottom panel of Figure~\ref{fig:skymap} we compare our data with g-levels measured by the IPS facility operated by the Institute for Space-Earth Environmental Research in Japan \cite<ISEE;>{1990SSRv...53..173K,2013PJAB...89...67T}.
The ISEE array observes sources as they cross the local meridian, and we have marked the location of this meridian at the time of the MWA observation (thus sources close to the dotted line, including those in the region of enhanced g-levels, will have been observed near-contemporaneously with the MWA observation).
One source inside the enhanced region (marked with a blue square in the top panel of Figure~\ref{fig:skymap}) was observed by ISEE with the g-level of 2.17 providing an independent verification.

The ISEE array consists of 3 stations separated by $\sim100$\,km from which the velocity of the scintillation pattern (and hence the solar wind) can be derived.
The source in our enhanced scintillation region has a velocity measurement of 453\,km s$^{-1}$, compared to equatorial measurements off the Western limb of the Sun in days prior/post of around 300 km s$^{-1}$.
Note, however, that no error is provided for these measurements, which means that only 2 stations were used.
This means that errors (which are typically $\sim$10\%) may be larger than usual, and therefore this measurement should not be given undue weight.

These ISEE data also drive the UCSD model \cite{1998JGR...10312049J,2020FrASS...7...76J}.
The all-sky v-maps (velocity) and g-maps generated by this model show no sign of the CME.
However, the velocity is consistently close to 300 km s$^{-1}$, at the times of both our reference and target observation and several days either side.
We assume this velocity for the background solar wind throughout the rest of this paper.

\subsection{MWA IPS Power Spectra}
As well as the change in scintillation level, IPS timeseries encode further information about the radio sources and scattering medium.
Most notably, a higher transverse velocity of the scattering screen will cause the scintillation signature to shift towards a higher frequency \cite{1990MNRAS.244..691M}.
\begin{figure}
    \centering
    \includegraphics[width=0.45\textwidth]{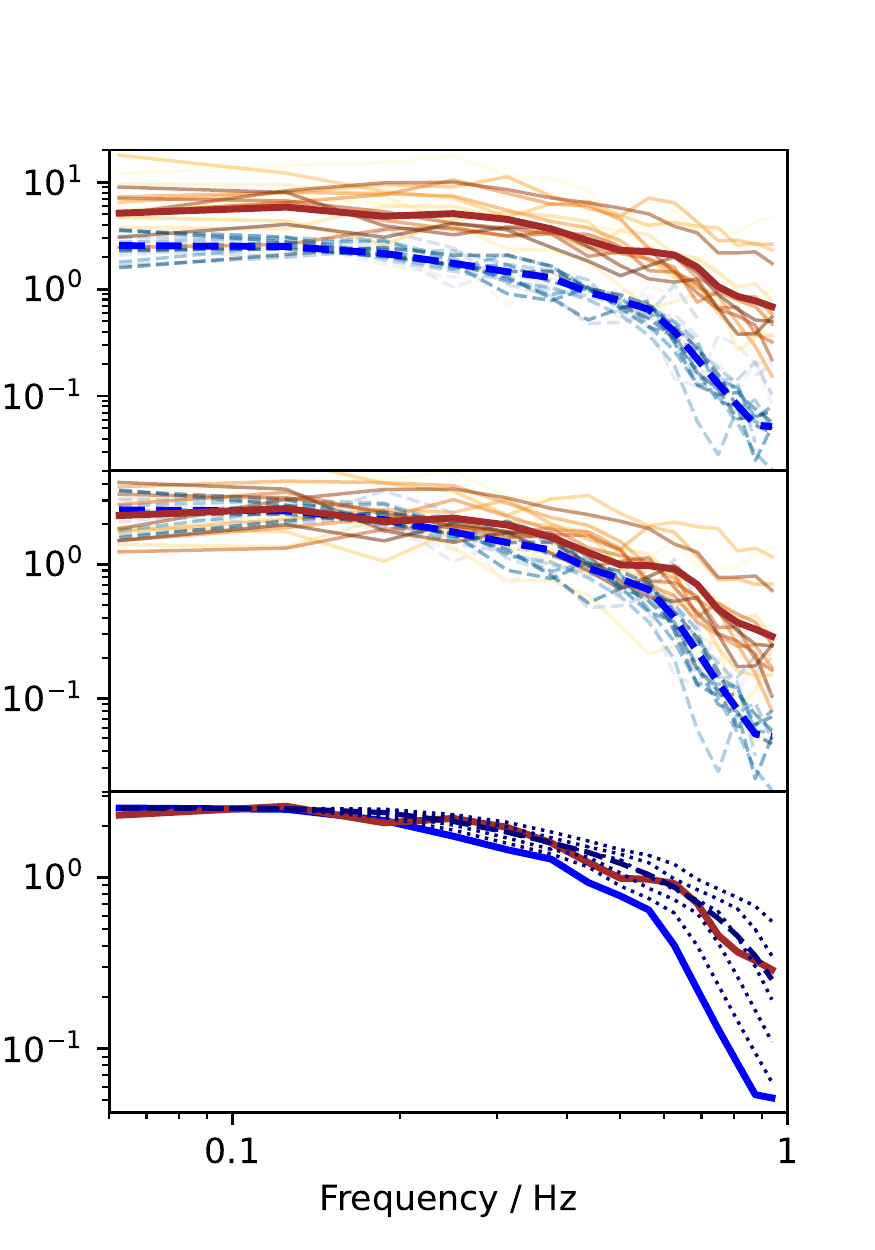} \,
    \caption{\label{fig:ps_multi}
    Top panel: power spectra for each source in the enhanced g region for the reference observation (in shades of blue), and target observation (in shades of brown).
    Darker shades are for sources with a higher S/N.
    Thick blue and brown lines are the average reference and target power spectrum respectively.
    Middle panel: same as top panel but target power spectra have been scaled by $g^{-2}$.
    Bottom panel: Average power spectra from 2nd panel.
    The Grey dotted lines are a set of reference power spectra scaled in frequency by factors of 1--1.6.
    Black dashed line is the average of the grey dotted lines.}
\end{figure}
Figure.~\ref{fig:ps_multi} shows the power spectra corresponding to the 576\,s timeseries, both in the target and reference observation, for those sources within the g=1.5 contour.
Only 15 sources with S/N$>10$ and scintillation index $>0.1$ in the reference observation have been selected.

The power spectra were constructed as follows: for each source, for both target and reference observations, timeseries for the maximum-brightness pixel for each source were extracted from each dataset, along with an off-source timeseries for a nearby pixel to provide a noise-level reference.
Power spectra were then formed for each timeseries, and the off-source power spectra (all of which were checked to be consistent with white noise) were subtracted from the corresponding on-source power spectra.
To permit comparison between different sources, all power spectra were normalised by the power spectrum from the reference observation in the range 0.25--0.5\,Hz (thus, by construction, the reference power spectra all lie on top of each other).
Note that variance due to ionospheric scintillation is typically two orders of magnitude weaker, and largely restricted to frequencies below 0.1\,Hz \cite{2022PASA...39...36W}, and so ionospheric effects are not expected to be visible.

The top panel shows both target and reference power spectra exactly as described above.
All show the shape which is characteristic of weak scintillation (flat at low frequencies, followed by a power-law drop off above the ``Fresnel Knee'').
The variance has typically dropped by an order of magnitude by the Nyquist frequency of 1\,Hz, but has not yet reached zero, so some high-frequency information has been lost.
Nonetheless, there is sufficient information to make some simple observations.

It is clear that the target power spectra differ considerably from one to the next, however there is no obvious spatial trend to this variability.
There also appears to be higher power at higher frequencies for the target power spectra.
To demonstrate this more clearly, in the middle panel we have scaled the target power spectra by a factor of $g^{-2}$, so the remaining difference between the power spectra cannot be explained by the overall change in the scintillation amplitude.

As shown in the lower panel, if we take the average reference power spectrum, scale it in frequency by a range of factors between 1.1 and 1.6 (5 different scalings are plotted as dotted lines), and take the average (the bold dashed line) we end up with power spectrum resembling the target power spectrum.
A physical interpretation for this is given in the following section.

\section{Analysis}
\label{sec:analysis}
The estimation of 3D CME structure from 2D images requires care at longer solar elongations due to the complex geometry and underlying physics \cite<e.g.>{2009SSRv..147...31H}.
Changes in the CME morphology, or deviation from simple models, can easily lead to errors \cite<e.g.>{2010ApJ...715.1524W}.
Here, we restrict ourselves to simple geometrical arguments, and basic scintillation physics to provide some approximate constraints on the CME kinematics.
\begin{figure}
    \centering
    \includegraphics[width=0.45\textwidth]{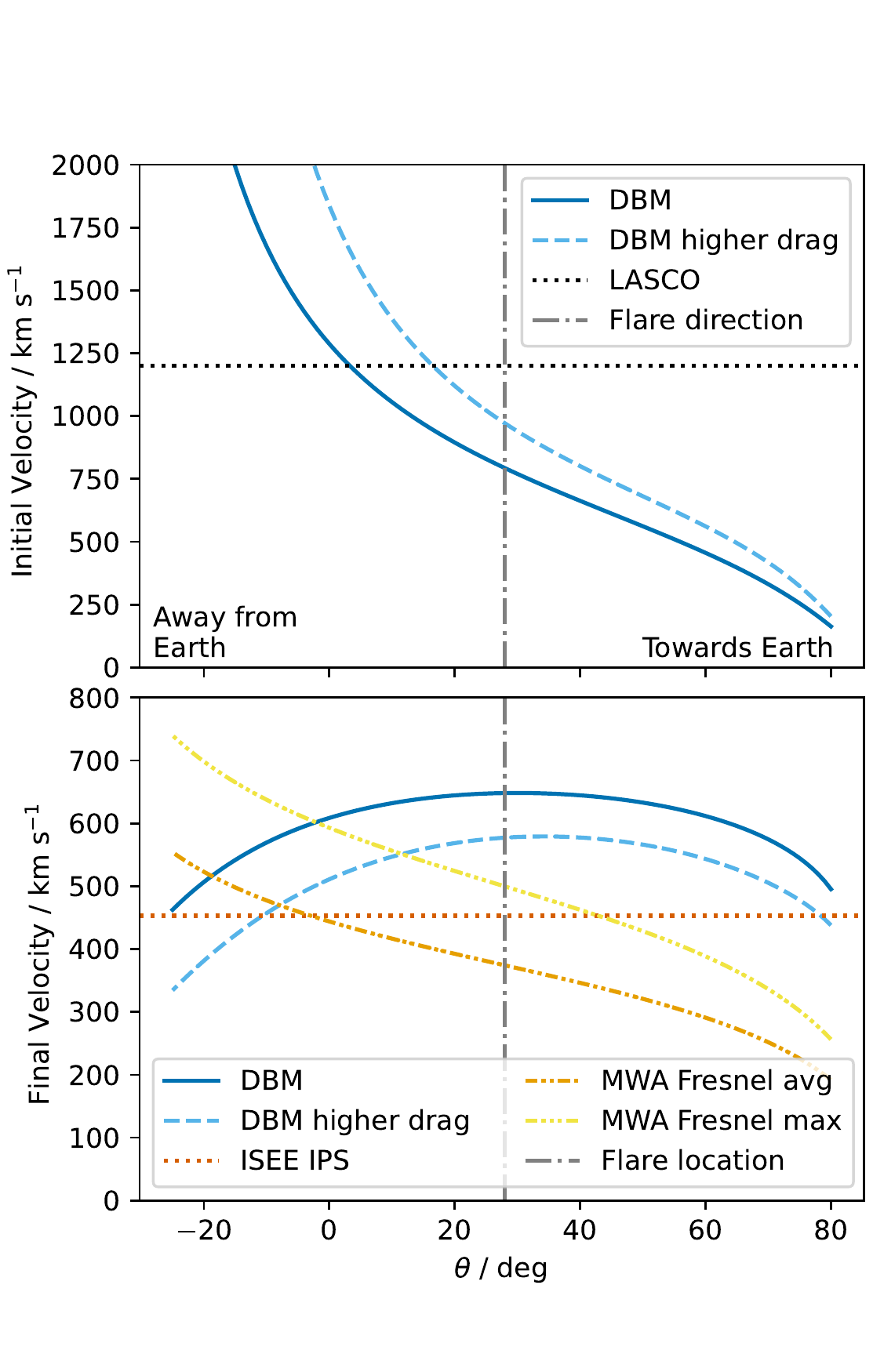} \,
    \caption{\label{fig:velocity}
    Top panel: initial velocity, projected into the plane of the sky, required to carry a CME from the LASCO-observed location to the MWA-observed elongation under the assumptions of the drag-based model (c.f. Figure~\ref{fig:lasco}).
    Results are shown assuming the standard value of $\gamma$ (=$10^{-8}$\,km$^{-1}$; solid line) and double that value (dashed line). 
    Dotted line shows observed initial velocity.
    Lower panel: blue lines show the final velocity implied by the initial velocities, again projected into the plane of the sky (i.e. component normal to the line of sight at an elongation of 37.5$^\circ$). Dotted red line shows velocity measured by multi-station IPS. Orange and yellow lines show the velocity implied by the mean (1.3) and max (1.6) frequency factors.}
\end{figure}

The location of the H$\alpha$ solar flare gives a strong indication of the origin of the CME.
\citeA{2008ApJ...673.1174Y} found that although CMEs associated with flares launch on a radial trajectory on average, there is a distribution of position angle differences with a standard deviation of 18$^\circ$.
Thus, the discrepancy between a radial trajectory from the flare ($\phi\approx10^\circ$) and the LASCO-observed trajectory ($\phi=22^\circ$) is unremarkable, and there will be a similar uncertainty in the angle to the plane of sky, $\theta$.

The main finding of our MWA observations is the location of the enhanced scattering, the leading edge of which lies at an elongation of approximately 37.5$^\circ$.
We now use the `Drag-based Model' (DBM) of \citeA{2013SoPh..285..295V} to confirm that this feature can be identified with CME detected previously in LASCO data.
In the DBM, the kinematics of the CME are based purely on the background solar wind speed ($w$, which we assume to be 300 km s$^{-1}$, see Sect.~\ref{sec:isee}) and a constant drag parameter, $\gamma$.

Under these assumptions, for a given, $w$, $\gamma$ and $\theta$, there is a unique initial velocity which will result in the CME traversing the direct path from an initial location close to the Sun (taken to be the point on the trajectory with $\epsilon=2.56^\circ$; see Sect.~\ref{sec:ancillary}), to the final location on the MWA-observed line of sight.
These implied initial velocities are plotted in the top panel of Figure~\ref{fig:velocity}, for a range of $\theta$, for the prescribed value of $\gamma$ for a strong, bright CME, and also for a slightly higher drag. 
The velocities have been projected into the plane of the sky to permit comparison with the coronagraph measurements.
There is good agreement of these velocities with the LASCO launch velocity estimate of 1200\,km\,s$^{-1}$ (see Sect.~\ref{sec:ancillary}) particularly for CME trajectories slightly closer to the plane of sky than suggested by the flare location.

A secondary finding is higher-than-ambient solar wind velocity in the heliosphere transient.
This is suggested by the multi-station IPS measurement made by ISEE (see Sect.~\ref{sec:isee} and Figure~\ref{fig:skymap}); though as noted, this is rather weak evidence since only two stations were used and the errors may be large.

The excess higher-frequency power in the MWA scintillation is also suggestive of a higher screen velocity, since the scintillation timescale ($\tau_\textrm{scint}$) is given by the Fresnel scale crossing time; i.e.
\begin{equation}
    \label{eqn:scint_time}
    \tau_\textrm{scint} = r_F/v ; \hspace{3ex} r_F=\sqrt{\frac{\lambda D}{2\pi}} ,
\end{equation}
where $v$ is the velocity and $r_F$ is the Fresnel scale, $\lambda$ is the observing wavelength, and $D$ is the distance from the observer \cite<e.g.>{1993ppl..conf..151N}.

The single-station IPS velocity measurement technique of \citeA{1990MNRAS.244..691M} introduced the method of measuring the shift in scintillation frequency and mapping this to a change in velocity.
Below, we extend this idea slightly by noting that Equation~\ref{eqn:scint_time} implies that a change in distance will also change the scintillation frequency.
Thus, we aim to find which effective distance (and therefore which value of $\theta$) gives the best consistency between our observed power spectra and other information on the likely velocity of the CME at the time of the IPS observation.

First we consider undisturbed solar wind which we presume is prevailing in the reference observation.
In this case the scintillation signal will be dominated by the wind close to the point of closest approach, due to the steep drop in solar wind density as it expands away from the Sun \cite<e.g.>{1976JGR....81.4797C}.
Consequently, the reference power spectra are a weighted average of scintillation along the line of sight, but with the weighting function peaked at the piercepoint $D=\cos\left(\epsilon\right)$\,AU \cite<see, e.g.>[Figure~1]{1991JGR....96.1717R} and $v=300$\,km s$^{-1}$ (see Sect.~\ref{sec:ancillary}).

However, when a heliospheric transient such as a CME shock is traversing the line of sight, its density may exceed that at the piercepoint, thereby shifting the weighted average distance away from the piercepoint.
Therefore, one interpretation of the bottom panel of Figure~\ref{fig:ps_multi} is that $v/r_F$ within the CME is between 1$\times$ and 1.6$\times$ higher than in the reference observation, averaged along and over different lines of sight.
The detailed IPS study of a very energetic CME by \citeA{2010SoPh..265..137M} is instructive in interpreting this: when individual sources were observed over several hours as the full CME structure crosses the line of sight, power-spectrum velocity measurements ranged from CME speed to closer to ambient.

In the lower panel of Figure~\ref{fig:velocity}, we plot the final velocity implied by the initial velocities in the upper panel and the DBM model.
As before, we project these velocities into the plane of the sky, this time for the line of sight of an observation offset by 37.5$^\circ$ from the Sun, to facilitate comparison with the IPS-derived velocities (This means that the velocities peak at $\theta=37.5^\circ$).

We also plot `Fresnel velocities': velocities implied by the observed increase in the frequency of the scintillation between reference and target observations.
Equation~\ref{eqn:sinerule} gives a monotonic relationship between $\theta$ and $D$, and thereby $r_F$ for our fixed value of $\epsilon$, allowing us to solve for $v$ in the target observation.

These IPS-derived velocities again favour a line of sight closer to the plane of sky ($\theta=0$) than the solar flare location suggests.
The average Fresnel velocity and multi-station velocity are in mild tension with the DBM according to this simple analysis.
However, as noted by \citeA{2021ApJ...922...73T}, this is to be expected since the CME will only occupy part of the line of sight.
These measurements are a weighted average over the entire line of sight and so will be regressed towards the ambient solar wind speed.

\section{Discussion}
\label{sec:discussion}
As we have shown, the region of enhanced interplanetary scintillation can be identified with the LASCO-detected CME.
The plane-of-sky velocity measured in the coronagraph images will carry the CME to the elongation observed with the MWA under reasonable assumptions.
Trajectories slightly closer to the plane of sky and/or slightly lower drag are favoured compared to the radial trajectory implied by the solar flare location.

Our velocity estimates based on the average power spectrum are broadly consistent with the near-contemp-oraneous multi-station IPS measurement by ISEE, and can be made consistent with the kinematics implied by the location of the CME and the DBM propagation model.
Likewise, the velocity analysis favours trajectories closer to the plane of sky and/or slightly lower drag.

However, this interpretation is not unique; for example, if the drag is higher closer to the Sun (and lower further away) then a more radial trajectory is consistent with our data.
MWA IPS observations are ongoing, and in order to confirm the value of these observations, it will be necessary to assemble a larger sample of CME observations.

\subsection{Future Work}
In the current work we report on a case where the CME was detected only in a single observation, which leaves some ambiguity as to the northern extent of the CME.
Since the MWA can be digitally steered to any point on the sky, in the future it will be possible to ascertain the extent of the CME more definitively, by extending our g-map using multiple pointings.

There are numerous possible approaches to determining CME speeds directly from our IPS data.
The simple power spectrum approach demonstrated here is consistent with our other data; in particular the comparison with the multi-station IPS measurement (which is independent of the Fresnel scale) appears to rule out an Earth-bound trajectory.
Secondly, once the individual MWA IPS sources are better understood, it will be possible to do more sophisticated power spectrum fitting to probe the turbulence properties, and the detailed morphology of a CME \cite<e.g.>{2021MNRAS.508.1314C}.
Thirdly, \citeA{2018MNRAS.473.2965M} have shown that double radio sources (a common morphology) also allow the measurement of the speed of the scintillation pattern, in an approach analogous to multi-station IPS.
Finally, since the MWA is not limited to a daily observing cadence, it is also possible for us to measure the plane of sky motion of a CME over time.
For example, had we observed our CME twice 2 hours apart, there would have been a shift in position of approximately 2$^\circ$, which would be easily measurable.

CME morphology in 3D has previously been modelled directly from IPS observations \cite{2003JGRA..108.1220T,2010SoPh..265..159T}, and IPS data has also been used to select the best candidate from an ensemble of MHD models \cite{2021EP&S...73....9I}.
Such approaches should be directly applicable to our data.

\citeA{Chhetri:2022} have recently demonstrated that the ASKAP telescope, which is co-located with the MWA, and is similarly characterised by a wide field of view, is capable of making IPS observations. 
ASKAP's higher operating frequency (700\,MHz--1800\,MHz) makes it more suited to observing closer to the Sun, so that it should be possible to observe a CME at any distance from $\sim5^\circ$ outwards with the MWA and/or ASKAP.
Utilising one or both these instruments, it will be possible to track a CME over a period of up to 12 hours, with thousands of sources all around the Sun being measured on an hourly cadence.

With a growing number of IPS-capable observatories world-wide, the heliosphere can be observed around the clock with a suite of instruments providing complementary information.
\citeA{Jackson:2022} have recently shown that data from multiple IPS observatories can be synthesised together to reconstruct the inner heliosphere, using tomographic techniques; thus a priority will be to test how well MWA IPS data can drive such reconstructions \cite<or MHD simulations; see e.g.>{2015SpWea..13..104J}.

Finally, this new capability has synergies with attempts to remotely sense the magnetic field orientation of a CME via Faraday rotation \cite<see>[for a recent review]{2022FrASS...941866K}.
\citeA{2016ApJ...831..208H} stress the importance of tracers of CME density (for which IPS g-levels, like white-light coronagraph measurements, can act as a proxy). 
By providing detailed CME plane-of-sky morphology, at greater distance from the Sun, IPS increases the sky area over which Faraday rotation observations of a CME can be made.
We note that our MWA IPS observations are also amenable to full polarimetric analysis \cite<see e.g.>[for a discussion of the polarimetric capabilities of the MWA]{2017PASA...34...40L}, and that MWA observations can be used to generate an exquisite map of the ionospheric electron density \cite{2015GeoRL..42.3707L,2017MNRAS.471.3974J}, a major contaminant of Faraday rotation observations \cite{2012RaSc...47.0K08O}.

\section{Open Research}


\footnotesize
The CACTus CME catalogue \cite{2009ApJ...691.1222R} used for the initial cross-match described in Sect.~\ref{sec:xmatch} is available online (at \url{https://www.sidc.be/cactus/})

Sunpy \cite{2020ApJ...890...68S} was used for coordinate conversions, as well as for downloading the GOES X-ray flux data as described in Sect.~\ref{sec:ancillary}.
We used Sunpy version 4.0.2 \cite{Mumford:2022}.
Sunpy was also used to retrieve Coronagraph images from \url{helioviewer.org}, including the one shown in the left panel of Figure~\ref{fig:lasco}.
At the time, helioviewer was running release 3.4.0 (\url{https://github.com/Helioviewer-Project/api/releases/tag/3.4.0}).

Data from the ISEE IPS array is available in near-real time, with archival data going back many years.
At the time of writing the ISEE data used in this paper can be found at the following URL: \url{https://stsw1.isee.nagoya-u.ac.jp/vlist/rt/nagoya.2016}.
Reconstructions of velocity and g-level fields from the same data can be found at \url{ips.ucsd.edu} under the section ``Archival 3D Imagery''.
The ``fisheye'' images use approximately the same coordinate system as used in this paper.

MWA data is available from the Australian Virtual Observatory \url{https://asvo.mwatelescope.org/}.
At the time of writing the observations used in this paper are public, and can be identified by their GPS start times (1147479952 and 1147393800) which serve as unique identifiers of these observations within the MWA archive.
All the IPS observations described in \cite{2019PASA...36....2M} are also archived, under project code \texttt{D0011}.
\newpage
\acknowledgments
JM and PIM would like to thank Iver Cairns (USyd) and Colin Lonsdale (MIT, Haystack) for organising and participating in the workshop which initiated this work.
This scientific work makes use of the Murchison Radio-astronomy Observatory, operated by CSIRO.
We acknowledge the Wajarri Yamatji people as the traditional owners of the Observatory site.
Support for the operation of the MWA is provided by the Australian Government (NCRIS), under a contract to Curtin University administered by Astronomy Australia Limited.
We acknowledge the Pawsey Supercomputing Centre which is supported by the Western Australian and Australian Governments.
JM acknowledges support from AFOSR grants FA2386-17-1-4126 and FA9550-18-1-0473.

\bibliography{refs.bib}
\end{document}